\documentclass[9pt]{article}
\usepackage{amssymb,latexsym,amsmath}
\usepackage{xcolor}
\usepackage{bm}

\usepackage{hyperref}

\begin{document}

\newtheorem{theorem}{Theorem}[section]

\newtheorem{proposition}[theorem]{Proposition}

\newtheorem{lemma}[theorem]{Lemma}

\newtheorem{corollary}[theorem]{Corollary}

\newtheorem{definition}[theorem]{Definition}

\newtheorem{remark}[theorem]{Remark}

\newtheorem{exempl}{Example}[section]

\newenvironment{exemplu}{\begin{exempl}  \em}{\hfill $\surd$

\end{exempl}}

\newcommand{\ea}{\mbox{{\bf a}}}
\newcommand{\eu}{\mbox{{\bf u}}}
\newcommand{\ep}{\mbox{{\bf p}}}
\newcommand{\ed}{\mbox{{\bf d}}}
\newcommand{\eD}{\mbox{{\bf D}}}
\newcommand{\eK}{\mathbb{K}}
\newcommand{\eL}{\mathbb{L}}
\newcommand{\eB}{\mathbb{B}}
\newcommand{\ueu}{\underline{\eu}}
\newcommand{\ueo}{\overline{u}}
\newcommand{\oeu}{\overline{\eu}}
\newcommand{\ew}{\mbox{{\bf w}}}
\newcommand{\ef}{\mbox{{\bf f}}}
\newcommand{\eF}{\mbox{{\bf F}}}
\newcommand{\eC}{\mbox{{\bf C}}}
\newcommand{\en}{\mbox{{\bf n}}}
\newcommand{\eT}{\mbox{{\bf T}}}
\newcommand{\eV}{\mbox{{\bf V}}}
\newcommand{\eU}{\mbox{{\bf U}}}
\newcommand{\ev}{\mbox{{\bf v}}}
\newcommand{\eve}{\mbox{{\bf e}}}
\newcommand{\uev}{\underline{\ev}}
\newcommand{\eY}{\mbox{{\bf Y}}}
\newcommand{\eP}{\mbox{{\bf P}}}
\newcommand{\eS}{\mbox{{\bf S}}}
\newcommand{\eJ}{\mbox{{\bf J}}}
\newcommand{\leb}{{\cal L}^{n}}
\newcommand{\eI}{{\cal I}}
\newcommand{\eE}{{\cal E}}
\newcommand{\hen}{{\cal H}^{n-1}}
\newcommand{\eBV}{\mbox{{\bf BV}}}
\newcommand{\eA}{\mbox{{\bf A}}}
\newcommand{\eSBV}{\mbox{{\bf SBV}}}
\newcommand{\eBD}{\mbox{{\bf BD}}}
\newcommand{\eSBD}{\mbox{{\bf SBD}}}
\newcommand{\ecs}{\mbox{{\bf X}}}
\newcommand{\eg}{\mbox{{\bf g}}}
\newcommand{\paromega}{\partial \Omega}
\newcommand{\gau}{\Gamma_{u}}
\newcommand{\gaf}{\Gamma_{f}}
\newcommand{\sig}{{\bf \sigma}}
\newcommand{\gac}{\Gamma_{\mbox{{\bf c}}}}
\newcommand{\deu}{\dot{\eu}}
\newcommand{\dueu}{\underline{\deu}}
\newcommand{\dev}{\dot{\ev}}
\newcommand{\duev}{\underline{\dev}}
\newcommand{\weak}{\rightharpoonup}
\newcommand{\weakdown}{\rightharpoondown}
\renewcommand{\contentsname}{ }

\title{\textbf{Symplectic and variational formulations\\
               of compressible and incompressible Navier-Stokes equation}}
\author{\textbf{G. de Saxc\'e} \\
Univ. Lille, CNRS, Centrale Lille, UMR 9013 – LaMcube \\
 Laboratoire de m\'ecanique multiphysique multi\'echelle, \\
 F-59000, Lille, France
\footnote{
Email address for correspondence: gery.de-saxce@univ-lille.fr}}

\maketitle

\begin{abstract}
In a previous paper, we proposed a symplectic version of Brezis-Ekeland-Nayroles principle based on the concepts of Hamiltonian inclusions and symplectic polar functions. We illustrated it by application to the standard plasticity in small deformations. The object of this work is to generalize the previous formalism to dissipative media in large deformations and Eulerian description. This aim is reached in three steps. Firstly, we develop a Lagrangian formalism for the reversible media based on the calculus of variation by jet theory. Next, we propose a corresponding Hamiltonian formalism for such media. Finally, we deduce from it a symplectic minimum principle for dissipative media and we show how to obtain a minimum principle for unstationary compressible and incompressible Navier-Stokes equation.
\end{abstract}



\section{Introduction}


\textit{This paper was written out to mark the occasion of the 200$^{\,th}$ birthday of Navier's works that spearheaded the Navier-Stokes equation. } 

\vspace{0.2cm}

Among the tools of the differential geometry, one of the most used is the geometry of the Riemannian spaces, equipped with a symmetric 2-covariant tensor, the metric
$$ g (dz, dz') = dz^T  K\, d z'
$$
allowing to measure lengths, but there is another interesting geometry, the one of the symplectic space, equipped with an antisymmetric 2-covariant tensor
$$ \omega (dz, dz') = dz^T  J\, dz'
$$ 
allowing to measure areas (\cite{Libermann Marle 1987}, \cite{Souriau 1997b}). One of the origins of the symplectic geometry is the dynamics of reversible systems. We are working with the phase space of which the elements are of the form
$$z  =\left[ \begin{array} {c}
                       \xi  \\
                       \eta\\
                    \end {array} \right] 
$$
where $\xi$ are the degrees of freedom and $\eta$ are the corresponding momenta. It is equipped with the canonical symplectic form 
$$\omega (d z, d z') =
                    \left[ d\xi , d\eta \right]
                    \left[ \begin{array} {cc}
                       0 & I   \\
                     - I  & 0  \\
                    \end {array} \right]\,
                    \left[ \begin{array} {c}
                       d \xi'  \\
                       d \eta' \\
                    \end {array} \right] 
$$
where  $I$ is the identity operator.  Introducing the Hamiltonian vector field (or symplectic gradient)
$$ \dot{z} = X_H = J \cdot \nabla_{z}  H (t, z )
$$
restitues the canonical equations
$$ \dot{\xi }   =   \nabla_{\eta } H, \qquad 
 \dot{\eta } = - \nabla_{\xi } H     
$$
Considering the lagrangian $L$ associated to the Hamiltonian $H$, this formulation can be recast in integral form thanks to

\textbf{Hamilton's variational principle:} \textit{the natural evolution of the systems minimizes}
 \textit{the action} 
  $$\alpha \left[\xi\right]= \int^{T}_0 L (\xi, \dot{\xi}, t) \,\mbox{ d} t $$  
\textit{among the paths} $t \mapsto \xi (t)$ \textit{satisfying}
$$\xi (0) = \xi_0, \quad 
    \xi (T) = \xi_T 
$$ 
It is worth to remark that the application of this principle is strictly limited to the reversible systems but unfortunately \textit{it fails} for dissipative ones. 
For dissipative systems, several authors proposed unified frameworks inspired from both overmentioned geometries. We quickly review now these theoretical frameworks: 
\begin{itemize}
\item The GENERIC systems (General Equation for Non-Equilibrium Reversible-Irreversible Coupling) were introduced by Grmla and \"Ottinger (\cite{Grmela 1997}, \cite{Ottinger 1997}). They combine the Hamiltonian formulation and the Onsager one, according to the evolution law
$$ \dot{z} = J \cdot \nabla_{z } H (z ) 
                        + K   \cdot \nabla_{z } S (z)
$$
where Onsager term is built from a symmetric and  positive-definite operator $K$ and an
entropy-like function $S$. A variational formulation of GENERIC can be found in \cite{Manh Hong Duong 2013}.
\item The Port-Hamiltonian systems were introduced by Brockett \cite{Brockett 1977} and van der Schaft \cite{van der Schaft 1984}
$$ \dot{z} = (J - R)\cdot \nabla_{z} H (z )
$$  
where the symmetric and positive-definite operator $ R$ modelizes the resistive effects (because of the minus sign).                   
\item The rate-independent systems proposed by Mielke and Theil \cite{mielketh99}, Mielke \cite{mielke} and developped with applications in Mielke and Roub\'{\i}\v{c}ek \cite{MR06b}, are based on two fundamental conditions:
\begin{itemize}
\item The stability condition: 
$ \nabla_{\xi} E (\xi) \cdot w + \Phi (w) \geq  0, \; \forall w $ 
\item The power balance: 
$ \nabla_{\xi} E (\xi) \cdot \dot{\xi} + \Phi (\dot{\xi} ) = 0 $ 
\end{itemize}
where $ E $ in the energy functional and $\Phi$ is a 1-homogeneous dissipation potential depending on the velocity.
\item The Hamiltonian inclusions  were proposed by Buliga \cite{bham},
\begin{equation}
 \dot{z} = J \cdot \nabla_{z } H (z ) +  J \cdot \nabla_{\dot{z} } \Phi (\dot{z} ) 
\label{dot(z) = J nabla_z H (z) + nabla_(dot(z)) Phi (dot(z))}
\end{equation}
where $ \Phi$ is a convex dissipation potential. Because of the first term, It is clearly related to the two former formalisms (GENERIC and Port-Hamiltonian systems) but it is also inspired from the rate-independent systems because of the dissipation potential. Nevertheless it is important to remark that $\Phi$ is not necessarily 1-homogeneous or even homogeneous. 
\end{itemize}

This Hamiltonian inclusion framework is compatible with the theory of internal variables and generalized standard materials applied successfully  to the modelling of the material behaviour in plasticity, viscoplasticity and damage \cite{Halphen 1975}. As such constitutive laws are non smooth, the non differentiability of the dissipation potential is no longer required and the second term in the expression (\ref{dot(z) = J nabla_z H (z) + nabla_(dot(z)) Phi (dot(z))}) is a non unique generalized gradient belonging to a set called the symplectic subdifferential \cite{bham}. The constitutive law is multivoque and the suitable mathematical tool is the convex analysis. As such non smooth laws are not necessary to build a variational principle for Navier-Stokes equation, we continue in this paper considering the more regular case where the dissipative law is univoque and the dissipation potential is differentiable but its convexity property remains essential.

In our opinion, the main advantage of Hamiltonian inclusion framework is that it leads naturally to a variational formulation with a minimum principle. Following Brezis and Ekeland \cite{Brezis Ekeland 1976} and Nayroles \cite{Nayroles 1976},
 Buliga and the author proposed in \cite{SBEN} a symplectic version of the Brezis-Ekeland-Nayroles principle (in short SBEN) of which the functional is built from three functions, the symplectic form (for the dynamics), the dissipation potential (for the irreversible behaviour) and the Hamiltonian (for the reversible behavior) through the Hamiltonian vector field.

Our main goal was to propose new numerical methods downstream for plasticity and viscoplasticity of structures in small deformations using the finite element method, more precisely to compute the whole loading history in contrast to now widely used incremental methods but that suffer from drawbacks with respect to stability and convergence criteria. We verified the feasability of these new numerical methods  in statics for thick tubes \cite{Cao 2020} and plates \cite{Cao 2021a} and in dynamics \cite{Cao 2021b}. Also we proposed a non incremental variational principle for brittle fracture using mainly the concept of driving force and with an enlightening discussion of crack stability criteria of the literature \cite{de Saxce 2022}.

The next step is extending this formalism in large deformation. For the solids, the suitable  approach is working with the material or Lagrangian representation. In \cite{Cao 2022}, we applied the SBEN principle to the plasticity with the additive decomposition of the Green-Lagrange tensor and the multiplicative decomposition of the deformation gradient.

To avoid any confusion, it must be pointed out that, although the SBEN approach uses tools of symplectic geometry, its application to Navier-Stokes equation has nothing to do with the formalism of Hamiltonian fluid dynamics \cite{Salmon 1988} which applies only to non dissipative fluids.
The problem of identifying a variational principle for the motion of a viscous fluid does not admit a unique solution. The pionnering works are due to Helmholtz \cite{Helmholtz 1869} (based on a minimum dissipation principle) and Rayleigh \cite{Rayleigh 1913}, both considering restrictive hypotheses on the boundary conditions and flow features. A more general result was proposed by Millikan \cite{Millikan 1929} and Bateman \cite{Bateman 1929}. Onsager discussed the validity of this principle and introduced a description of the production of entropy in terms of thermodynamic forces and fluxes \cite{Onsager 1931}. Glansdorff and Prigogine \cite{Glansdorff Prigogine 1964} derived the approach of the fluid flow variational that was re-elaborated by Gyamati \cite{Gyamati 1970}. Lebon and Lambermont proposed in \cite{Lebon 1973} a general variational principle for unstationary dissipative thermodynamics processes. Lebon et al. \cite{Lebon 1976} applied it to problems with unsteady laminar viscous flows.
Sciubba proposed a variational principle for Navier-Stokes equation based on the concept of exergy in \cite{Sciubba 1991} and \cite{Sciubba 2005}. Besides, the last reference presents a comprehensive survey of previous works on the topics. To deduce Navier-Stokes equation in a thermodynamical framework, Hamilton's variational principle was modified by  Fukagawa and Fujitani \cite{Fukagawa 2012}  by introducing a nonholonomic constraint and Lagrange multipliers. Applying the optimal control theory, they discuss their approach in connection with the Hamiltonian formulation. 
Likewise, Gay-Balmaz and Yoshimura \cite{Gay-Balmaz 2017b} proposed a Lagrangian variational description of the Navier-Stokes-Fourier system by introducing new dual variables of which the time rates are the production of entropy and the temperature. 
The variational formulation in spatial representation is obtained using the Euler–Poincaré reduction with advected parameters proposed by Holm et al. \cite{Holm 1998}. In \cite{Raza 2006}, Razafindralandy and Hamdouni proposed for incompressible Navier-Stokes equation to use a bi-Lagrangian but this approach leads, by adding adjoint variables, to double the number of unknown fields and equations.
The nearest approach to the present one is Ghoussoub and Moameni work on the application of anti-selfdual Lagrangian to the resolution of the Navier-Stokes evolutions (\cite{Ghoussoub 2005}, \cite{Ghoussoub 2007}, \cite{Ghoussoub 2009}). Using convex analysis and functional analysis tools and under suitable assumptions, they prove the existence of solutions as minimizers of a given functional for particular boundary conditons but their approach is quite general and may apply to many other situations.  In \cite{Parker 2022}, Parker and Schneider use a variational method for finding periodic orbits in the incompressible Navier–Stokes equations but without using their dynamical or symplectic structure. 

Our variational approach to Navier-Stokes is different because developped from a geometric point of view  as in Gay-Balmaz and Yoshimura work \cite{Gay-Balmaz 2017b}. It leads to construct a problem of minimizing a functional equivalent to compressible Navier-Stokes equation. This equation is covariant in the sense that it includes the gravitation and satisfies Galileo's principle of relativity.  The limit case of the incompressible flow is deduced by well-known variational techniques. The SBEN principle is a geometric method to built variational principles for dissipative systems, not only for solids but also for fluids. The main technical difficulties is that, unlike the applications to solid mechanics obtained customarily working with the material representation, the modelization of fluids requires to work in the spatial or Eulerian representation. 

Our goal is reached in three steps. In Section \ref{Section - Lagrangian formalism for a reversible continuum}, we develop for the non dissipative systems a Lagrangian formalism based on the calculus of variation performed on the jet space of order one. 
In Section \ref{Section - Hamiltonian formalism and canonical equations for a reversible continuum}, we propose a corresponding Hamiltonian formalism. 
In Section \ref{Section - Symplectic Brezis-Ekeland-Nayroles principle for dissipative continua}, we deduce a symplectic variational principle of minimum for the dissipative systems in Eulerian representation. We apply it to the compressible Navier-Stokes equation in Section \ref{Section - SBEN principle for compressible Navier-Stokes equation} and to the limit case of incompressible Navier-Stokes equation in Section \ref{Section - SBEN principle for incompressible Navier-Stokes equation}.

\section{Lagrangian formalism for a reversible continuum}

\label{Section - Lagrangian formalism for a reversible continuum}

We are working in the space-time, set of events $X = (t, x)$ where $x$ is the position at time $t$. In contrast to the usual representation of the motion of a continuum by a map $x = f (t, x_0)$ where $x_0$ are the material  coordinates of the particle, our point of view, inspired from Hilbert-Einstein action in General Relativity \cite{GR}, is to represent it by  $x_0 = \kappa (t, x)  = \kappa_t (x) = \kappa (X)$.
As usual, the deformation gradient is denoted 
\begin{equation}
F = \nabla_{x_0} x
\label{F = nabla_(x_0) x}
\end{equation}
Let $\Omega$ be a bounded open subset of the space-time corresponding to the  motion of the continuum. The equations of balance of the linear momentum and the energy of a reversible continuum are deduced from a space-time action 
$$ \alpha \left[x_0\right] = \int_\Omega \mathcal{L} \left( X, x_0, \nabla_X x_0\right) \mbox{d}^4 X
$$
of Hamilton's principle, using a special form of the calculus of variation by replacing the original field $x_0$ by its first jet prolongation $j^1 x_0$, a map from $\Omega$ into the jet bundle $J^1 (\Omega, \mathbb{R} ^3)$ such that $j^1 x_0 (X) = (X, x_0 (X),   \nabla_X x_0 (X))$, that leads to perform variations not only on the field and its derivatives but also on the variable $X$. To explicit them, we consider a new parameterization given by a regular map $X = \psi\ (Y)$ of class $C^1$ and we perform the variation of the function $\psi$, the new variable being $Y$. After calculating the variation of the action, we will consider the particular case where the function $\psi$ is the identity of $\Omega$.
Hence we start with
\begin{equation}
   \alpha \left[X, x_0\right] = \int_{\Omega'} 
                               \mathcal{L} \left( \psi(Y), x_0, 
                                       \nabla_Y x_0 \cdot \nabla_X Y\right)\
                                       \det \left(  \nabla_Y X  \right) 
                           d^4 Y
\label{alpha (X, x_0) =}
\end{equation}
where $\Omega' = \psi^{-1} (\Omega)$ and the variables of the functional are now both $X$ and $x_0$. For more details on the jet spaces and the calculus of variation on such spaces, the reader is referred to (\cite{Saunders}, \cite{Aldaya}, \cite{Edelen}, \cite{Mangiarotti}, \cite{AffineMechBook}). For  sake of easiness, we introduce
\begin{equation}
     f = - \nabla_{x_0} \mathcal{L}, \quad
     h = \nabla_X \mathcal{L}, \quad
     P = \nabla_{\nabla_X x_0} \mathcal{L}, \quad
     T = P \cdot \nabla_X x_0 - \mathcal{L} \, I
\label{def f & h & P & T}     
\end{equation}
The technical details of the variation are given in Appendix A. Taking the variation with respect to $x_0$ and $X$, we obtain Euler-Lagrange equations of variation:
\begin{equation}
    \nabla_X \cdot P +   f = 0, \qquad
   \nabla_X \cdot T + h  = 0\ .
\label{variation equation} 
\end{equation} 
The first equation leads to a non linear partial derivative system which can be used to determine the unknown field $x_0$. The last one gives extra conservation conditions obtained in the spirit of Noether's theorem.

In order to satisfy Galileo's principle of relativity, we consider the Lagrangian
\begin{equation}
  \mathcal{L} = \rho \,\left(\frac{1}{2}\,\parallel v \parallel^2 + A \cdot v - \phi - e_{int} (x_0, C)\right)
\label{L = T - U} 
\end{equation} 
where $v = - F \cdot (\partial x_0 / \partial t)$ is the spatial velocity, $A, \phi$ are the vector and scalar potentials of the Galilean gravitation \cite{AffineMechBook}, $e_{int}$ is the specific internal energy depending on the right Cauchy strains 
\begin{equation}
    C = F^T \cdot F
\label{C = F^T cdot F}
\end{equation}
and 
\begin{equation}
    \rho = \frac{\rho_0 (x_0)}{\det (F)} 
\label{rho = rho_0 (x_0) / det (F)} 
\end{equation} 
is  the mass density satisfying the equation of balance of mass
\begin{equation}
   \frac{\partial \rho}{\partial t} + \nabla \cdot (\rho \, v) = 0
\label{balance of mass}
\end{equation}
Introducing the linear momentum 
\begin{equation}
   \pi = \nabla_v \mathcal{L} = \rho\,(v + A)\ ,
\label{generalized linear momentum pi = rho (v + A)} 
\end{equation}
the Hamiltonian density
\begin{equation}
   \mathcal{H} =  \pi \cdot v - \mathcal {L} = \frac{1}{2\,\rho}\, \parallel \pi - \rho\,A \parallel^2 + \rho\,(\phi + e_{int})\ .
\label{H = (1/2) rho norm(pi - rho A)^ 2 + rho (phi + e_int)} 
\end{equation}
and the reversible stresses
\begin{equation}
  \sigma_R  =   2 \rho\,F\cdot \nabla _C \, e_{int}\cdot F^T
\label{sigma = - 2 rho F (partial L / partial  C) F^T} 
\end{equation}
In the sequel, the gradient $\nabla_x$ with respect to the position will be denoted $\nabla$ as usual.
As shown in \cite{AffineMechBook}, it holds
\begin{equation}
   T =\left[ \begin{array} {cc}
                         \mathcal{H}                & - \pi             \\
                         \mathcal{H} v - \sigma_R v & \sigma_R - v \otimes \pi \\
                 \end {array} \right],\qquad
   h =  \left[ \begin{array} {c}
                    \rho \left(   \frac{\partial A}{\partial t}\cdot v -   \frac{\partial \phi}{\partial t}  \right)                   \\
                    \rho \, ( (\nabla  A) \cdot v - \nabla \phi )      \\
                 \end {array} \right]
\label{standard T} 
\end{equation}
$T$ is the analogous in Galilean mechanics of the energy-momentum tensor in relativistic one. Then the equation (\ref{variation equation}) of variation of the action with respect to $X = (t, x)$ allows we recover the balance equations of  the energy 
$$ \frac{\partial \mathcal{H}}{\partial t} 
    + \nabla \cdot \left( \mathcal{H} \,  v - \sigma_R \cdot  v \right)
    =  \rho \left(\frac{\partial \phi}{\partial t} - \frac{\partial A}{\partial t}\cdot v\right)
$$
and the balance of linear momentum
\begin{equation}
    - \frac{\partial \pi} {\partial t} + \nabla \cdot (\sigma_R - v \otimes \pi)
    + \rho \, ( (\nabla  A) \cdot v - \nabla \phi ) = 0
\label{balance of linear momentum by calculus of variation intermediate expression} 
\end{equation}
After classical simplifications owing to the balance of mass and given in Appendix B, it is reduced to 
\begin{equation}
    - \rho\,\frac{D v}{D t} + \nabla \cdot \sigma_R  + \rho\,\left(g - 2\,\Omega \times v \right) = 0
\label{balace of linear momentum}
\end{equation}
where occurs the material derivative $D / Dt = \partial / \partial t + v \cdot \nabla $, the gravity $ g$ and Coriolis' vector $ \Omega$ defined by
\begin{equation}
    g = - \nabla  \phi - \frac{\partial A}{\partial t}, \qquad
    \Omega = \frac{1}{2} \, \nabla \times A
\label{def g & Omega}
\end{equation}
 For a barotropic fluid, $ \sigma_R = - p \, I$ where $p$ is the pressure  then the balance of linear momentum takes the form of Euler's equations
\begin{equation}
 - \rho\,\frac{D v}{D t}  - \nabla p  + \rho\,\left(g - 2\,\Omega \times v \right) = 0
\label{balance of linear momentum for barotropic fluids} 
\end{equation}

It is worth  to remark these two equations of balance are covariant with respect to Galileo's principle of relativity. The last term is Coriolis' force useful for the applications to environmental fluids (atmosphere, oceans, Earth's interior).

\section{Hamiltonian formalism and canonical equations for a reversible continuum}

\label{Section - Hamiltonian formalism and canonical equations for a reversible continuum}

Let $\Omega_t$ the set of positions occupied by the material particles of the continuum at time $t$.
As $v$ are the components of a 1-contravariant tensor and $\mathcal{H} =  \pi \cdot v - \mathcal {L}$ is a density, $\pi$ are the components of a 1-covariant and antisymmetric 3-contravariant tensor. Then the field $x \mapsto \pi (t, x)$ defined on $\Omega_t$ is a section of the fiber bundle $\bigwedge^3 (T \, \Omega_t) \otimes T^* \Omega_t $ of which the coordinates $(x, \pi)$ in local charts are taken as canonical variables. 
The total energy at time $t$ is
$$ H \left[ x_0, \pi \right]   = \int_{\Omega_t} \mathcal{H} \left( x, x_0, \nabla x_0, \pi \right) d^3 x
$$
where, for sake of easiness, the dependence with respect to time $t$ is no longer explicitly expressed in the sequel.

We claim that the motion of the continuum is described by the canonical equations
$$\zeta = \left(\frac{d x}{d t}, \frac{\partial \pi}{\partial t}\right) = \left(v, \frac{\partial \pi}{\partial t}\right) = X_H
$$
obtained by the calculus of the Hamiltonian vector field $X_H$ for the canonical symplectic form 
\begin{equation}
 \omega (\zeta, \zeta') = \int_{\Omega_t} \left( 
\frac{d x}{d t}\cdot \frac{\partial \pi'}{\partial t}
- \frac{\partial \pi}{\partial t} \cdot \frac{d x'}{d t}
\right) \mbox{d}^3 x
\label{omega (zeta, zeta'}
\end{equation}

As $H$ is a functional, $X_H$ is a variational derivative that can be calculated by the jet theory as in the previous section. Hence we consider a new parameterization $x = \psi\ (y)$ of class $C^1$ and we perform the variation of the function $\psi$, the new variable being $y$. After calculating the variation of the functional, we will consider the particular case where the function $\psi$ is the identity of $\Omega_t$.
Hence we start with
\begin{equation}
   H \left[x, x_0, \pi' \right] = \int_{\Omega'_t} 
                               \mathcal{H} \left( \psi(y), x_0, 
                                       \nabla_y x_0 \cdot \nabla y,
                                       \det \left( \nabla y\right)\, (\nabla y)^T \cdot \pi'\right)\
                                       \det \left(  \nabla_y x \right) 
                           d^3 y \ ,
\label{H (y, x_O, pi') =}
\end{equation}
where $\Omega'_t = \psi^{-1} (\Omega_t)$ and $\pi'$ are the  components of the linear momentum in the coordinates $y$. A calculus similar to the one of the previous section and detailled in Appendix C leads to the following expression of the canonical equations
$$   \frac{d x}{d t} =  \nabla_\pi \mathcal{H}
$$
$$ \frac{\partial \pi}{\partial t} =  -  \nabla \mathcal{H}
$$
$$  -  \nabla \cdot \left[
     \nabla_{\nabla  x_0} \, \mathcal{H}
   \cdot \nabla  x_0
   - \left(\mathcal{H} -  \nabla_\pi \mathcal{H} \cdot \pi \right) \, I
   + \nabla_\pi \mathcal{H}\otimes \pi
   \right]
$$
where the extra terms of the jet theory are given by the last line. 

Let us determine these equations for the Hamiltonian (\ref{H = (1/2) rho norm(pi - rho A)^ 2 + rho (phi + e_int)}) corresponding to the Lagrangian (\ref{L = T - U}). We have
\begin{equation}
     \nabla_\pi \mathcal{H} = \frac{\pi}{\rho} - A\ . 
\label{V = pi/rho - A} 
\end{equation}
$$ \nabla_{\nabla  x_0} \, \mathcal{H} = - 2\, \nabla_{x_0} x \cdot \nabla_C\mathcal{H} \cdot C
$$
which, taking into account (\ref{F = nabla_(x_0) x}), (\ref{C = F^T cdot F}) and (\ref{rho = rho_0 (x_0) / det (F)}), reduces to:
$$ \nabla_{\nabla  x_0} \, \mathcal{H}  = - \left(   \dfrac{1}{2\,\rho}\, \parallel \pi \parallel^2
                    - \dfrac{\rho}{2}\, \parallel A \parallel^2
                    - \rho\,(\phi + e_{int}) 
             \right)\, \nabla_{x_0} x
           - 2\,\rho \, \nabla_{x_0} x \cdot \nabla_C e_{int} \cdot C
$$
Taking into account (\ref{sigma = - 2 rho F (partial L / partial  C) F^T}) and (\ref{V = pi/rho - A}), we obtain
\begin{equation}   
        \nabla_{\nabla  x_0} \, \mathcal{H}
   \cdot \nabla  x_0
   - \left(\mathcal{H} -  \nabla_\pi \mathcal{H} \cdot \pi \right) \, I
   + \nabla_\pi \mathcal{H}\otimes \pi
       = - \sigma_R + v \otimes \pi\ .
\label{T^* = - sigma_R + v pi^T} 
\end{equation}
Besides
\begin{equation}
   -  \nabla \mathcal{H} = \rho\,\left( (\nabla A) \cdot \left(\frac{\pi}{\rho} - A\right) - \nabla \phi\right) 
         = \rho\,\left( (\nabla \, A) \cdot v - \nabla \phi\right) \ .
\label{h^* = rho ((grad A) v - grad phi)} 
\end{equation}

For the Lagrangian (\ref{L = T - U}), we obtain
\begin{equation} 
\frac{d x}{d t} = \frac{\pi}{\rho} - A , \qquad
     - \frac{\partial \pi} {\partial t} + \nabla \cdot (\sigma_R - v \otimes \pi)
    + \rho \, ( (\nabla  A) \cdot  v - \nabla \phi ) = 0
\label{canonical equations of a reversible continuum}
\end{equation}
where we can recognize the definition (\ref{generalized linear momentum pi = rho (v + A)}) of the linear momentum and the equation (\ref{balance of linear momentum by calculus of variation intermediate expression}) of balance of the linear momentum \cite{AffineMechBook}.

\section{Symplectic Brezis-Ekeland-Nayroles principle for dissipative continua}
\label{Section - Symplectic Brezis-Ekeland-Nayroles principle for dissipative continua}

To build a minimum principle for such continua as a viscous fluid, we adapt the scheme proposed in \cite{SBEN}. The key-idea is a decomposition of $\zeta$ into reversible and irreversible parts
$$ \zeta = \zeta_R + \zeta_I, \qquad \zeta_R = X_H, \qquad \zeta_I = \zeta - X_H
$$
We are interested in the material continua that are modelized by a differentiable and convex potential of dissipation $\Phi$. We claim that the law of dissipative yielding is given by 
\begin{equation}
   \zeta_I = X_\Phi
\label{zeta_I = X_Phi}
\end{equation}
where the Hamiltonian vector field $X_\Phi$ is such that
$$ \forall \zeta',\qquad 
    \omega (X_\Phi, \zeta')
    = \lim_{\epsilon \to 0} \frac{1}{\epsilon} 
      (\Phi (\zeta + \epsilon \, \zeta')
      - \Phi (\zeta))
$$ 
Because of the convexity property \cite{bham}
$$  \forall \zeta', \quad \Phi (\zeta + \zeta') - \Phi (\zeta) \geq \omega (X_\Phi, \zeta') 
$$
Next, we define the symplectic Fenchel polar (or conjugate) function $\Phi^{*\omega}$ by
\begin{equation}
 \Phi^{*\omega} (\zeta_I) = \sup_{\zeta} \;
    (\omega (\zeta_I, \zeta) - \Phi (\zeta))
\label{def symplectic Fenchel polar}
\end{equation}
As superior envelop of affine functions, it satifies the symplectic Fenchel inequality
\begin{equation}
\forall  \zeta', \forall  \zeta'_I, \qquad
 \Phi(\zeta') + \Phi^{*\omega} (\zeta'_I) - \omega( \zeta'_I, \zeta')  \geq 0
 \label{symplectic Fenchel inequality}
 \end{equation}
In particular, when the dynamical dissipative constitutive law (\ref{zeta_I = X_Phi}) is satisfied, it can be proved the equality is reached \cite{SBEN}
 \begin{equation}
 \zeta_I = X_\Phi \qquad \Leftrightarrow \qquad
 \Phi(\zeta) + \Phi^{*\omega} (\zeta_I) - \omega( \zeta_I, \zeta)  = 0
 \label{dynamical dissibative constitutive law}
 \end{equation}

We are interested during the interval $\left[ 0, T \right]$ by the evolution paths $t \mapsto (\kappa_t, \zeta)$ which are admissible in the sense that they satisfy the boundary conditions and the initial conditions of the considered problem (that we shall suppose well posed).
Following pionnering works by Brezis, Ekeland and Nayroles in 1976,
 Buliga and the author proposed in \cite{SBEN} a symplectic version of the Brezis-Ekeland-Nayroles principle (in short SBEN) of which the functional is built from three functions, the symplectic form (for the dynamics), the dissipation potential (for the irreversible behavior) and the Hamiltonian (for the reversible behavior) through the Hamiltonian vector field:
 
\textit{SBEN principle: the natural evolution path} $t \mapsto (\kappa_t, \zeta)$ \textit{minimizes the functional}
\begin{equation}
 \Pi [\kappa, \zeta]  =
    \int^T_0 \lbrace \Phi (\zeta) + \Phi^{*\omega} (\zeta - X_H)
               - \omega (\zeta - X_H, \zeta )  \rbrace\ \mbox{d}t 
\label{Pi (kappa, zeta)}
\end{equation}
\textit{among all the admissible evolution paths, and the minimum is zero.}

\vspace{0.15cm}

The idea is that the functional is non negative because of (\ref{symplectic Fenchel inequality}) and vanishes if and only if the  dynamical dissipative constitutive law (\ref{zeta_I = X_Phi}) is satisfied almost eveywhere because of (\ref{dynamical dissibative constitutive law}). 

\section{SBEN principle for compressible Navier-Stokes equation}

\label{Section - SBEN principle for compressible Navier-Stokes equation}

With the notations of the previous section, the canonical equations (\ref{canonical equations of a reversible continuum}) lead to 
$$ \zeta_I = \zeta - X_H = (v_I, \pi_I)
$$
with 
$$  v_I = v - \frac{\pi}{\rho} + A , \quad
     \pi_I = \frac{\partial \pi} {\partial t} - \nabla \cdot (\sigma_R - v \otimes \pi)
    - \rho \, ( (\nabla  A) \, v - \nabla \phi ) 
$$
With the same classical simplifications used to transform (\ref{balance of linear momentum by calculus of variation intermediate expression}) into (\ref{balance of linear momentum for barotropic fluids}), we have for a barotropic fluid
\begin{equation}
\pi_I =   \rho\,\frac{D v}{D t}  + \nabla p  - \rho\,\left(g - 2\,\Omega \times v \right) 
\label{pi_I for barotropic fluids} 
\end{equation}
To apply the very general formalism of the previous section to Navier-Stokes equation, we need two additional hypotheses. The first one claims that 
 the convex smooth potential $ \Phi$ depends explicitly only on $v$ ($\partial \pi / \partial t$ is ignorable):
\begin{equation}
  \Phi (\zeta) = \varphi (v)
\label{phi (zeta) = varphi (v)} 
\end{equation}
that reads, taking into account (\ref{omega (zeta, zeta'}) and (\ref{phi (zeta) = varphi (v)}):
$$ \Phi^{*\omega} (\zeta_I) = 
     \sup_\zeta \; \left\lbrace \int_{\Omega_t} \left(v_I\cdot \frac{\partial \pi}{\partial t} - v\cdot \pi_I \right) d^3 x
      - \varphi (v) \right\rbrace
$$
where $\zeta = (v, \partial \pi / \partial t)$. Hence, it holds:
$$ \Phi^{*\omega} (\zeta_I)  = 
     \sup_v \; \left\lbrace \int_{\Omega_t} v\cdot ( - \pi_I  ) \,  d^3 x
      - \varphi (v)  \right\rbrace
     + \sup_{\tiny{\partial \pi / \partial t}} \; \int_{\Omega_t} v_I\cdot \frac{\partial \pi}{\partial t} \,  d^3 x
$$
where the value of the last term is zero if $v_I = 0$ and $+\infty$ otherwise.
Then the symplectic Fenchel polar function has a finite value 
$$    \Phi^{*\omega} (\zeta_I) = \Phi^{*\omega} (v_I, \pi_I) = \varphi^* (- \pi_I) 
$$
if $v_I = 0$, that is (\ref{generalized linear momentum pi = rho (v + A)}), where $\varphi^*$ is the classical Fenchel polar function of $\varphi$ \cite{Ekeland Temam 1999}.
As in the SBEN principle the minimum value of the functional is zero then finite, we suppose in the sequel that (\ref{generalized linear momentum pi = rho (v + A)}) is \textit{a priori} satisfied. Moreover, owing to (\ref{omega (zeta, zeta'}), the last term in the functional (\ref{Pi (kappa, zeta)}) becomes 
$$ - \omega (\zeta - X_H , \zeta )
   = \int_{\Omega_t} \left( 
\pi_I \cdot v
- v_I \cdot \frac{\partial \pi}{\partial t}
\right) \mbox{d}^3 x
   = \int_{\Omega_t}  
\pi_I \cdot v \,
 \mbox{d}^3 x
$$
Then the SBEN functional becomes
$$  \Pi [\kappa, \zeta]  =
    \int^T_0 \lbrace \varphi (v) + \varphi^{*} (- \pi_I)
              + \int_{\Omega_t}  
\pi_I \cdot v \,
 \mbox{d}^3 x  \rbrace\ \mbox{d}t 
$$
Owing to (\ref{pi_I for barotropic fluids}), the functional (\ref{Pi (kappa, zeta)}) can be recast, that leads to

\vspace{0.15cm}

\textit{SBEN principle for compressible Navier-Stokes equation: \\
the natural evolution path} $t \mapsto (\kappa_t, v)$ \textit{minimizes the functional}
\begin{eqnarray}
 \Pi [\kappa, v]  & = &
    \int^T_0 \lbrace \varphi (v) 
                              + \varphi^{*} \left( - \rho\,\frac{D v}{D t}  - \nabla p  + \rho\,\left(g - 2\,\Omega \times v \right) \right) \nonumber\\
         &   &     + \int_{\Omega_t}  
\left[ \rho\,\frac{D v}{D t}  + \nabla p  - \rho\, g  \right] \cdot v \,
 \mbox{d}^3 x  \rbrace\ \mbox{d}t 
\label{Pi (kappa, v)}
\end{eqnarray}
\textit{among all the admissible evolution paths, and the minimum is zero.}

\vspace{0.15cm}

It is worth to remark that 
$  \Phi (\zeta)  + \Phi^{*\omega} (\zeta_I)  
    = \varphi (v) + \varphi^* (- \pi_I) $
is an anti-selfdual Lagrangian, a tool proposed by Ghoussoub \cite{Ghoussoub 2007} to study among others the solutions of Navier-Stokes equation. This reveals the symplectic origin of the structure of such self-antidual Lagrangians. On the other hand, while the last term of  Ghoussoub's functional is depending on specific boundary value problems, ours is general and does not depend on the boundary conditions that are only considered as constraints in the minimization.

The second additional hypothesis claims that $\varphi$ depends on $v$ through its symmetric gradient $D = \mathcal{D} (v) = \nabla_s v = 1 / 2 \, (\nabla v + (\nabla v)^T)$ and is quadratic with respect to $v$ of the form
$$ \varphi (v) = \int_{\Omega_t} W (\mathcal{D} (v))\,d^3 x
    = \int_{\Omega_t}   \mu\,\left[ Tr (D^2) - \frac{1}{3}\,(Tr (D))^2\right]  \,\mbox{d}^3 x
$$
then the viscous part of the stress tensor is traceless (Stokes hypothesis)
\begin{equation}
   \sigma_I = \nabla_D W (\mathcal{D} (v)) =  2 \mu\, \left( D - \frac{1}{3}\,Tr (D)\, I\right)
\label{Stokes hypothesis}
\end{equation}
As $\varphi$ is a functional, its variational derivative is 
$$ \nabla_v \varphi (v)  = - \nabla \cdot \sigma_I
$$
Now, let us prove that the variational principle of minimum restitues Navier-Stokes equation. Indeed, if the minimum equal  to zero is reached, we have almost everywhere in $\left[ 0, T  \right] $
$$ \varphi (v) + \varphi^{*} (- \pi_I)
              + \int_{\Omega_t}  
\pi_I \cdot v \,
 \mbox{d}^3 x = 0
$$
that is equivalent to the dynamical dissipative law
$$ - \pi_I =  \nabla_v \varphi (v)    = - \nabla \cdot \sigma_I
$$
Owing to (\ref{pi_I for barotropic fluids}) and (\ref{Stokes hypothesis}), we recover Navier-Stokes equation
$$    \rho\,\frac{D v}{D t} = - \nabla p + \mu\,\triangle v + \frac{\mu}{3}\,\nabla \,(\nabla \cdot v)
                              + \rho\,\left(g - 2\,\Omega \times v \right)\ .
$$
For a flow, the term of the last line in the functional (\ref{Pi (kappa, v)}) is the sum of the velocity head, pressure head and elevation head losses due to dissipation during the interval from $0$ to $T$.  For the limit case of inviscid flows, the potential of dissipation $\varphi$ vanishes and its polar function $\varphi^*$ has a  finite value equal to zero if $\pi_I = 0$, \textit{i.e.} Euler's equations (\ref{balance of linear momentum for barotropic fluids}), then the SBEN principle claims that the total head loss is zero, that is the expression of Bernoulli's principle. It is worth to notice that in this limit case the SBEN principle does not degenerate into Hamilton's principle. 

Likewise an argument developped in Proposition $2$ of \cite{Brezis Ekeland 1976} for a simpler case where $\mathcal{D}$ is the gradient of a scalar field, we can show that, if $W$ is quadratic and if the velocity or the dissipative stress vector is null on the boundary, Fenchel polar function of $\varphi$ is:
$$ \varphi^* (f) = \int_{\Omega_t} W (\mathcal{D} (K^{-1} (f)))\,d^3 x\ .
$$
\label{thm if W is quadratic Fenchel conjugate of varphi is}
where the linear operator $K$ is define by
\begin{equation}
   f = K (v) = - \nabla \cdot \left( \nabla_D W (\mathcal{D}(v))\right)
\label{g = A (v) = (div ( (partial W / partial D) (D (v))))^T} 
\end{equation}
The proof is given in Appendix D.

Although we study dissipative systems, we do not use concepts of the Thermodynamics. The framework of thermodynamic forces and fluxes, introduced by Onsager \cite{Onsager 1931} and used in their variational formulation by Glansdorff and Prigogine \cite{Glansdorff Prigogine 1964},  Lebon and Lambermont \cite{Lebon 1973}, is not used explicitly in our formalism but it is not far behind. In fact, the dissipation  potential depends on $v$ through its symmetric gradient $D$ of which the components can be taken as thermodynamic fluxes, then the corresponding thermodynamic forces are the components of  dissipative stress tensor $\sigma_I$ given by (\ref{Stokes hypothesis}).

\section{SBEN principle for incompressible Navier-Stokes equation}

\label{Section - SBEN principle for incompressible Navier-Stokes equation}

For this limit case, $\nabla \cdot v = 0 $ and the pressure $p$ becomes a free variable independent of $\kappa$. Navier-Stokes equation is reduced to:
 $$    \rho\,\frac{D v}{D t} = - \nabla p + \mu\,\triangle v 
                              + \rho\,\left(g - 2\,\Omega \times v \right)\ .
$$ 
To obtain the corresponding SBEN principle, we proceed as follows. The internal energy is cancelled in (\ref{L = T - U}) and (\ref{H = (1/2) rho norm(pi - rho A)^ 2 + rho (phi + e_int)}).
The incompressibility condition is introduced as a constraint in the minimization. The pressure disappears of the functional and reappears as a Lagrange multiplier of this constraint in the equation characterizing the minimizers. Then we state

\vspace{0.15cm}

\textit{SBEN principle for incompressible Navier-Stokes equation: \\the natural evolution path} $t \mapsto (\kappa_t, v)$ \textit{minimizes the functional}
\begin{eqnarray}
 \Pi [\kappa, v]   = 
    \int^T_0 \left\lbrace \varphi (v) 
                              + \varphi^{*} \left( - \rho\,\frac{D v}{D t}   + \rho\,\left(g - 2\,\Omega \times v \right) \right) 
                 + \int_{\Omega_t}  
\rho\, \left( \frac{D v}{D t}  - g \right) \cdot v \,
 \mbox{d}^3 x  \right\rbrace\ \mbox{d}t \nonumber
\label{Pi (v)}
\end{eqnarray}
\textit{among all the admissible evolution paths such that $\nabla \cdot v = 0 $, and the minimum is zero.}

\vspace{0.15cm}

If the fluid is homogeneous, the mass density $\rho$ is constant and the functional no longer depends on the motion map $\kappa$.

\section{Conclusions and perspectives}

In this work, the emphasis was laid on the geometric aspects thanks to the use of differential geometry tools in order to give a strong foundation to the formulation of variational principles for the dynamics of dissipative continuous media in spatial or Eulerian representation.
The present variational approach covers a large class of problems including Navier-Stokes equations for compressible and incompressible flows.  The weak regularity of the potential of dissipation adopted in \cite{SBEN} allows to encompass nonsmooth dissipative constitutive laws such as the one of 
Bingham fluids. Also, it is worth to observe that the expression of the functional is independent of the boundary conditions that appear only as constraints of the minimization. Moreover, the functional is not convex with respect to the unknown fields but there is  (at least partial)  convexity,  that is favourable for the convergence of the minimization procedure. 

\vspace{0.3cm}

The present work paves the way to provide numerical approximations of the solutions. Many numerical schemes fail to correctly predict the expected conservation of mechanical quantities such as the energy, as the time increases. Geometric integrators appear to be robust for large time simulation because they are based on the underlying geometric structure of the equations. Among them, we can quote the symplectic integrators (\cite{Simo 1992}, \cite{Hairer 2002}, \cite{Raza 2018}, \cite{Di Stasio 2019}), the variational integrators (\cite{Marsden 1998}, \cite{Gay-Balmaz 2018}, \cite{Raza 2018}, \cite{Raza 2019}) and Dirac integrators \cite{Raza 2019}. 
Another idea is to develop symplectic integrators to respect the structure of the canonical equation of Section \ref{Section - Hamiltonian formalism and canonical equations for a reversible continuum}.   Another perspective is to construct variational schemes based on the Lagrangian of the SBEN principles proposed in Sections \ref{Section - SBEN principle for compressible Navier-Stokes equation} and \ref{Section - SBEN principle for incompressible Navier-Stokes equation}. Also we think that the SBEN principle could be a new tool in theory of turbulence to propose variational-based models for Large Eddy Scale simulations.  Moreover, when the convergence of the minimization algorithm is reached, the value of the functional of SBEN principle for the numerical solution can be interpreted as an  indicator of the error in constitutive law (or in dissipation) introduced by Ladev\`{e}ze \cite{Ladeveze 2005}.

\vspace{0.3cm}

Finally a minimum variational principle may constitute an interesting means to study the existence and smoothness of the solutions of Navier-Stokes equations in the spirit of Ghoussoub and Moameni papers  (\cite{Ghoussoub 2005}, \cite{Ghoussoub 2007}, \cite{Ghoussoub 2009}).

\vspace{0.3cm}

\textbf{Acknowledgements}

\vspace{0.3cm}

The author would like to thank Maurice Rossi and Djimedo Kondo for their comments and suggestions that allowed to improve the paper and the \textit{Agence Nationale de la Recherche} (ANR) for the financial support of the project \textit{BIpotentiels Généralisés pour le principe variationnel de  Brezis-Ekeland-Nayroles en mécanique} (BigBen) that has enabled to perform this work. 

\vspace{0.3cm}

\textbf{Author ORCID.}

\vspace{0.3cm}

G. de Saxc\'e http://orcid.org/0000-0002-0961-0513.




\vspace{0.5cm}

\textbf{Appendix A}

\vspace{0.3cm}

Owing to (\ref{def f & h & P & T}), the variation of the action (\ref{alpha (X, x_0) =})  reads:
$$ \delta \alpha  = \int_{\Omega'}\ 
                [\left( Tr\ \left(  P \cdot 
                              \delta\ \left( \nabla_Y x_0 \cdot \nabla_X Y \right)\right)
                                        - f \cdot \delta x_0 + h \cdot \delta X \right) \
                                              \det\ \left( \nabla_Y X \right) 
$$
\begin{equation}
                                 +  \mathcal{L}\ \delta\ \left(\det\ \left(\nabla_Y X  \right) \right)  
                ] \,\mbox{d}^4 Y\ .\qquad\qquad
\label{delta I} 
\end{equation} 
                                   
First of all, we calculate the variation of the field derivative in terms of the derivative of its variation:
$$ \delta \left(\nabla_Y x_0 \cdot \nabla_X Y \right)
     = \delta \left( \nabla_Y x_0  \right) \cdot \nabla_X Y
      + \nabla_Y x_0\cdot  \delta\ \left( \nabla_Y X \right)^{-1} $$
\begin{equation}
    \delta \left(\nabla_Y x_0 \cdot \nabla_X Y \right)
        = \nabla_Y (\delta x_0)\cdot \nabla_X Y  
         - \nabla_Y x_0 \cdot 
           \nabla_X Y  \cdot \nabla_Y (\delta X)\cdot \nabla_X Y \ .
\label{delta derivative} 
\end{equation}  
Incidently, it is worth noting that when $X = Y$:
$$ 
   \delta \left(\nabla_X x_0 \right)
        = \nabla_X (\delta x_0)
         - \nabla_X x_0 \cdot 
           \nabla_X (\delta X)\ .
$$
This formula shows that, unlike the usual rule used in the classical calculus of variation, the derivative symbols $ \nabla_X$ and $\delta$ may not be permuted in the present approach. Next, using the adjugate of a matrix  $ \mbox{adj} (M) = \det (M) \, M^{-1}
$, one has:
\begin{equation}
   \delta \left(\det\ \left(  \nabla_Y X \right) \right) 
   = Tr\ \left(\nabla_Y  (\delta X) \cdot  adj \left( \nabla_Y X \right)    
    \right) 
 \label{delta det} 
\end{equation} 
Introducing the expressions (\ref{delta derivative}) and (\ref{delta det}) into the variation of the action (\ref{delta I}) gives:
$$ \delta \alpha  = \int_{\Omega'}\ [ 
                                   Tr\ \left(  P\cdot  \nabla_Y (\delta x_0)\cdot  
                                                 \det\ \left(  \nabla_Y X\right)\
                                                 \nabla_X Y\right) 
                                                  -  \det\ \left(  \nabla_Y X \right)\ 
                                                       (f\cdot \delta x_0 - h\cdot \delta X) $$
\begin{equation}
                                   + Tr\ \left(  \mathcal{L}\  \nabla_Y (\delta X)\cdot 
                                                 adj \left(  \nabla_Y X \right)
                                                - P\cdot \nabla_X x_0 \cdot  \nabla_Y (\delta X)\cdot
                                                     \det\ \left(  \nabla_Y X\right)\
                                                 \nabla_X Y\right) 
                                              ]\
                           \mbox{d}^4 Y\ .
\label{delta I bis} 
\end{equation} 
Taking into account Cramer's rule and  using the definition (\ref{def f & h & P & T}),  
the variation of the action (\ref{delta I bis}) becomes:
$$ \delta \alpha  = \int_{\Omega'}\ [\ 
                                   Tr\ \left(  adj\ \left( \nabla_Y X  \right)\cdot
                                                P\cdot \nabla_Y (\delta x_0) \right) \                                                
                                                  -  \det\ \left( \nabla_Y X  \right)\ 
                                                  (f\cdot  \delta x_0 - h\,\delta X)
$$
\begin{equation}
                                   - Tr\ \left(  adj\ \left(  \nabla_Y X \right)\cdot T\cdot
                                             \nabla_Y (\delta X)\ \right) 
                                              ]\
                           \mbox{d}^4 Y\ .
\label{delta I ter} 
\end{equation} 
Integrating by part and taking into account the fact that the values of $x_0$ and $X$ are imposed on the boundary, the surface integrals vanish and we obtain:
$$
    \delta \alpha  =   \int_{\Omega'}\  (\ - \left[ \ 
                                   \nabla_Y \cdot  \left(  adj\ \left( \nabla_Y X \right)   \cdot  P \right)                                               
                                                + \det\ \left( \nabla_Y X \right)\, f \right] \cdot \delta x_0
$$
$$
                                 + \left[
                                    \nabla_Y \cdot  \left(  adj\ \left( \nabla_Y X  \right)\cdot  T \right)  
                                 + \det\ \left(\nabla_Y X  \right) \, h \right] \cdot \delta X             )\
                           \mbox{d}^4 Y\ ,
$$
Finally, considering the particular case where $X = Y$, the variational principle reads:
$$
    \delta \alpha  =  \int_\Omega\  (\ - \left[ \nabla_X \cdot  P +   f \right] \cdot \delta x_0 
                + \left[ \nabla_X  \cdot T   + h \right] \cdot \delta X
                                                   )\
                           \mbox{d}^4 X = 0\ .
$$
Then we obtain Euler-Lagrange equations of variation (\ref{variation equation}).

\vspace{0.5cm}
\textbf{Appendix B}

\vspace{0.3cm}

Developing the balance of linear momentum (\ref{balance of linear momentum by calculus of variation intermediate expression}), we have:
$$  -  \frac{\partial }{\partial t} (\rho\,(v + A)) + \nabla \cdot \sigma_R 
      - div\,(\rho v)\,(v + A) - \rho \,  \nabla (v + A) \cdot v
      + \rho \, ( (\nabla  A) \cdot v - \nabla \phi )  = 0\ ,
$$ 
or, expanding the first term:
$$ - \rho\ \frac{\partial }{\partial t} (v + A) 
   - \left( \frac{\partial \rho }{\partial t}+ div\,(\rho v)\right)\,(v + A) - \rho \, \nabla (v + A) \cdot v
   + \nabla \cdot \sigma_R + \rho \, ( (\nabla  A) \cdot v - \nabla \phi )  = 0\ .
$$
Taking into account the balance of mass (\ref{balance of mass}), it holds:
$$ - \rho\ \left(\frac{\partial v}{\partial t} + (v \cdot \nabla) v \right) +\nabla \cdot \sigma_R
   -  \rho \left( \nabla \phi + \frac{\partial A }{\partial t}\right)
    + \rho v \cdot  \left( \nabla A - (\nabla A)^T \right) = 0\ .
$$
that, owing to (\ref{def g & Omega}), leads to the new form of the balance of linear momentum (\ref{balace of linear momentum}).

\vspace{0.5cm}
\textbf{Appendix C}

\vspace{0.3cm}

For sake of easiness, we introduce 
$$   V        =   \nabla_\pi \mathcal{H},\qquad 
       h^*    =  - \nabla \mathcal{H},\qquad
       P^*_x =  \nabla_{\nabla x_0} \mathcal{H}
$$
The variation of the action (\ref{H (y, x_O, pi') =}) with respect to the canonical variables $x$ and $\pi'$ reads:
$$ \delta H  = \int_{\Omega'}\ 
    [\left( Tr\ 
            \left(  P^*_x\cdot  \delta\ \left( \nabla_y x_0 \cdot \nabla y  
                               \right)
            \right) +  V\cdot \delta \left(\det \left( \nabla y \right)\, \left( \nabla y\right)^T\pi' \right) - h^* \cdot \delta x 
     \right) 
$$
\begin{equation}
  \ \det \ \left(  \nabla_y x \right)    +  \mathcal{H}\ \delta\ \left(\det\ \left(\nabla_y x   \right)  
        \right)        ] \,d^3 y\ .
\label{delta H} 
\end{equation}                                
First of all, we calculate the variation of the field derivative:
\begin{equation}
   \delta \left(  \nabla_y x_0 \cdot  \nabla y\right)
        = - \nabla_y x_0 \cdot 
          \nabla y\cdot \nabla_y (\delta x)\cdot \nabla y \ .
\label{delta derivative H} 
\end{equation}  
Once again, using the adjugate of a matrix $ \mbox{adj} (M) = \det (M) \, M^{-1}
$, one has:
\begin{equation}
   \delta \left(\det\ \left(  \nabla y \right) \right) 
            = Tr\ \left( \nabla_y (\delta x)\cdot adj \left(\nabla_y x\right) \right) \ . 
\label{delta det H} 
\end{equation}  
Besides, the variation of the momentum  $\pi$ reads:
\begin{eqnarray}
 \delta \left(\det \left( \nabla y \right)\, \left( \nabla y \right)^T \cdot \pi' \right) 
 & = & - \left(\det \left( \nabla y\right) \right)^2 
    Tr \left( \nabla_y (\delta x)\cdot adj \left(\nabla_y x\right)\right) \cdot \left(\nabla y \right)^T \cdot \pi' \nonumber\\
 &   & + \det \left( \nabla y\right)\,\left(\nabla y\right)^T \cdot
  \left(\delta \pi' - \left(\nabla_y (\delta x)\right)^T \cdot \left(\nabla y\right)^T \cdot \pi' \right)\nonumber
\end{eqnarray}
Introducing the expressions (\ref{delta derivative H}), (\ref{delta det H}) and the latter one into the variation of the Hamiltonian (\ref{delta H}) and using Cramer's rule gives:
\begin{eqnarray}
 \delta H & = & \int_{\Omega'}\ [ 
 V \cdot \left(\left( \nabla y 
               \right)^T \cdot \delta \pi' 
         \right)
- (h^* \cdot \delta x)\, \det\ \left(\nabla_y x
                               \right)\nonumber\\
& & - Tr\
\lbrace ( 
    adj \left(\nabla_y x 
              \right) \cdot
          \left(P^*_x\cdot \nabla x_0 
                - \left(\mathcal{H} - V\cdot \left(\det\,\left(\nabla y 
                                                          \right)\,
                                                   \left(\nabla y 
                                                   \right)^T \cdot \pi'
                                             \right)
                  \right)\,I
          \right) \nonumber\\
& &       + \nabla y \cdot (V \otimes \pi') \cdot  \nabla y 
   )\cdot \nabla_y (\delta x)
\rbrace\ 
]\ d^3 y\ .\nonumber
\end{eqnarray}
Integrating by parts the last term with vanishing surface integrals, we obtain:
$$ \delta H  = \int_{\Omega'}\ [ V \cdot \left(\left(\nabla y
               \right)^T \cdot \delta \pi' \right)
- (h^* \cdot \delta x)\, \det\ \left(  \nabla_y x \right)$$
$$ + \nabla_y\cdot  
        \lbrace adj \left(\nabla_y x \right) \cdot 
              \left(P^*_x\cdot \nabla x_0
                - \left(\mathcal{H} - V\cdot \left(\det\,\left(\nabla y
                                                          \right)\,
                                                   \left(\nabla y
                                                   \right)^T \cdot \pi'
                                             \right)\right)\,I \right) 
$$
$$   + \nabla y \cdot (V \otimes \pi') \cdot  \nabla y 
              \rbrace \cdot \delta x]\ d^3 y\ .
$$
Considering the particular case where $x = y$ (then $\pi = \pi'$), the variational principle reads:
\begin{equation}
   \delta H  = \int_{\Omega}\ [ V \cdot \delta \pi - h^* \cdot \delta x + (\nabla \cdot T^*)\cdot \delta x] \ d^3 x\ ,
\label{delta H = int_(omega) (V cdot delta pi - h^ * cdot delta x + (div T^ *) delta x) d^3 x}
\end{equation}
introducing the $3 \times 3$ matrix:
\begin{equation}
   T^* = P^*_x\cdot \nabla x_0 
                - \left(\mathcal{H} - V\cdot \pi\right)\, I  + V\otimes \pi
\label{T^* = P^*_x (partial x_0 / partial x) - (H - V cdot pi) 1_(R^3) - V pi^T} 
\end{equation}

\vspace{0.5cm}
\textbf{Appendix D}

\vspace{0.3cm}

Fenchel polar function of $\varphi$ is \cite{Ekeland Temam 1999}:
\begin{equation}
   \varphi^* (f) = 
     \sup_v  \, \int_{\Omega_t} [f\cdot v - W (\mathcal{D}(v))]\,d^3 x 
\label{Fenchel conjugate of varphi} 
\end{equation}
Denoting $\mathcal{L}_\varphi$ the integrand in the above expression, the supremum is reached when Euler-Lagrange equation is satisfied:
$$ \nabla_v \mathcal{L}_\varphi =   \nabla \cdot \left( \nabla_{\nabla v} \mathcal{L}_\varphi\right)
$$
The right hand member defines a linear map $K$ from the velocity field $v$ onto a vector field $f$ of which the value is the internal force (by unit volume) given  by (\ref{g = A (v) = (div ( (partial W / partial D) (D (v))))^T}). If the associated boundary value problem is well posed, the map is regular and $ v = K^{-1} (f)$. Integrating by parts, it holds:
$$ \int_{\Omega_t} \nabla \cdot \left( \nabla_D W (\mathcal{D}(v))\cdot v\right)\,d^3 x = 
   \int_{\Omega_t} 
   \left[Tr\, 
         \left\lbrace( \nabla_D W (\mathcal{D}(v)) \cdot \mathcal{D}(v)
         \right\rbrace 
         - f\cdot v\right] 
   \,d^3 x\ .
$$
Applying the divergence formula to the first term of the right hand member and taking into account that the velocity or the dissipative stress vector $\sigma_I \cdot n$ vanishes on the boundary of $\Omega_t$ of unit normal $n$, the left hand member is null. Then Fenchel polar function (\ref{Fenchel conjugate of varphi}) becomes:
$$    \varphi^* (f) = 
      \int_{\Omega_t} 
     \left[ Tr \,  \left\lbrace \nabla_D W (\mathcal{D}(v)) \cdot \mathcal{D}(v)
             \right\rbrace 
               - W (\mathcal{D}(v))
      \right]\,d^3 x 
$$
Since $W$ is quadratic and $ v = K^{-1} (f)$ the proof is achieved. \cite{Batchelor59}

\end{document}